\documentclass[twocolumn,amsthm]{autart}    
                                     
\usepackage{graphicx}
\usepackage{dcolumn}
\usepackage{bm}

\usepackage{csquotes}
\usepackage{enumitem}
\usepackage{xcolor}
\usepackage{amsmath}
\usepackage{amssymb}

\usepackage{enumitem}

\renewcommand{\t}[1]{\mathrm{#1}}

\newcommand{\RRR}{{\mathbb{R} }}

\newcommand{\bs}[1]{\boldsymbol{#1}}

\newcommand{\bdq}{\bs{\dot{q}}}
\newcommand{\bdgamma}{\bs{\dot{\gamma}}}

\newcommand{\ww}{\bs{w}}

\newcommand{\XX}{\bs{X}}

\newcommand{\ggamma}{\bs{\gamma}}

\newcommand{\nnabla}{\bs{\nabla}}

\newcommand{\nnull}{\bs{0}}

\renewcommand{\d}{{\mathrm{d}}}
\newcommand{\td}[2]{\frac{\d #1}{\d #2}}
\newcommand{\pdx}[2]{\frac{\partial #1}{\partial #2}}

\newcommand{\tensors}[3]{{#1^{#2}}_{#3}}

\newcommand{\ie}{i.\,e.}

\newcommand{\eg}{e.\,g.}
\newcommand{\wrt}{w.\,r.\,t.\ }

\newtheorem{theorem}{Theorem}
\newtheorem*{theorem*}{Theorem}

\newtheorem{definition}{Definition}
\newtheorem*{definition*}{Definition}

\newtheorem*{lemma*}{Lemma}

\newtheorem{corollar}{Corollary}
\newtheorem*{corollary*}{Corollary}

\newtheorem*{proof*}{Proof}

\makeatletter
\renewcommand*\env@matrix[1][\arraystretch]{%
	\edef\arraystretch{#1}%
	\hskip -\arraycolsep
	\let\@ifnextchar\new@ifnextchar
	\array{*\c@MaxMatrixCols c}}
\makeatother

\begin{document}

\begin{frontmatter}

\title{One-dimensional solution families of nonlinear systems characterized by scalar functions on Riemannian manifolds\thanksref{footnoteinfo}} 

\thanks[footnoteinfo]{This paper was not presented at any IFAC 
meeting. \\ Corresponding author A.~Albu-Sch\"affer. Tel. +49-172-3072097}

\author[DLR,TUM]{Alin Albu-Sch\"affer}\ead{alin.albu-schaeffer@dlr.de}, 
\author[DLR]{Dominic Lakatos}\ead{dominic.lakatos@dlr.de},              
\author[Twente]{Stefano Stramigioli}\ead{'s.stramigioli@utwente.nl'}  

\address[DLR]{German Aerospace Center}                                               
\address[TUM]{Technical University of Munich}             
\address[Twente]{University of Twente}


\begin{abstract}                          
For the study of highly nonlinear, conservative dynamic systems, finding special periodic solutions which can be seen as generalization of the well-known normal modes of linear systems is very attractive. 
However, the study of low-dimensional invariant manifolds in the form of nonlinear normal modes is rather a niche topic, treated mainly in the context of structural mechanics for systems with Euclidean metrics, i.e., for point masses connected by nonlinear springs. Newest results emphasize, however, that a very rich structure of periodic and low-dimensional solutions exist also within nonlinear systems such as elastic multi-body systems encountered in the biomechanics of humans and animals or of humanoid and quadruped robots, which are characterized by a non-constant metric tensor. This paper discusses different generalizations of linear oscillation modes to nonlinear systems and proposes a definition of strict nonlinear normal modes, which matches most of the relevant properties of the linear modes. The main contributions are a theorem providing necessary and sufficient conditions for the existence of strict oscillation modes on systems endowed with a Riemannian metric and a potential field as well as a constructive example of designing such modes in the case of an elastic double pendulum. 
\end{abstract}

\end{frontmatter}

\section{Introduction}
\label{sec:introduction}
The evolution of many physical systems is modeled by nonlinear second-order ordinary differential equations (ODEs). Explicit solutions of such equations are known only for very specific cases of nonlinear ODEs. For the particular, standard case of \emph{energy-conservative linear} systems of second-order ODEs analytical solutions are determined by the underlying generalized eigenvalue problem. Each conjugate complex eigenvalue pair and its corresponding eigenvectors determines a family of solutions, which is known as a mode of the linear system. Such linear normal modes share the following common properties:
\begin{itemize}
	\item[i.] solutions corresponding to a mode are periodic;
	\item[ii.] motions corresponding to a mode are such that the time evolution of all dependent variables (and their time-derivatives) are determined by a single second-order differential equation;
	\item[iii.] motions of all dependent variables are functionally related (actually linearly related in the specific case of linear systems) to a single dependent variable, forming straight modal lines in configuration space; 
	\item[iv.] the straight modal lines of (iii) are energy independent, \ie, the system evolves along those lines for any initial velocity along the lines. 
\end{itemize}
Although qualitatively distinct, the above properties are simultaneously satisfied for energy-conservative, linear systems. This stands in strong contrast to the nonlinear system case, where these properties are not necessarily linked. 

The general class of conservative nonlinear systems considered in this paper is characterized by the metric field $g$ and the scalar function $f : \mathcal M\to\RRR$ defined on a manifold $\mathcal M$. That is, $\left(\mathcal M, g\right)$ is an $n$-dimensional Riemannian manifold, where the two-covariant metric tensor field ${g : \mathcal M\to T^\star_p\mathcal M\otimes T^\star_p\mathcal M}$ assigns a positive definite inner product $\left<\cdot, \cdot\right>$ in each tangent space $T_p\mathcal M$. As a consequence, it is possible to define an affine connection $\nnabla$ (the Levi-Civita connection), compatible to the metric, which means that $\nnabla_X g=0,\, \forall X 
\in T_p\mathcal M$. Then, by letting $q(t)$ be a trajectory of points in $\mathcal M$ (\ie, the dependent variables) parametrized by $t\in\RRR$ (\ie, the independent variable), and $\bdq \in T_{q(t)}\mathcal M$ the associated tangent vector field, the considered nonlinear system of second-order ODEs can be expressed as
\begin{align}
\nnabla_{\bdq}\bdq = \nnabla f \,.
\label{eq:differential-equations}
\end{align} 
Herein, $\nnabla f$ denotes the contravariant gradient vector associated to the covector $\d f$ and satisfying $\d f(\ww) = \left<\nnabla f, \ww\right>, \, \forall \ww \in T_p\mathcal M$.
In case of a mechanical system $q\in\mathcal M$ are configuration variables, $t$ is time, $\bdq$ a vector field of velocities, $g$ the inertia tensor, $f$ a potential function, and $\nnabla f$ the acceleration due to $f$.\\
Depending on how many of the above four properties of linear modes one would like to preserve, different generalizations for nonlinear systems can be defined. Periodicity is the most general, yet most unspecific property, solutions of nonlinear systems might be required to obey defining a mode. Such a definition of nonlinear modes has been considered in \cite{Rand1971}, \cite{Rand1974}, and \cite{Zhang2005}. Demanding in addition to periodicity that  motions corresponding to a nonlinear mode are driven by a single second-order differential equation, leads to the definition of modes proposed by \emph{Shaw and Pierre} \cite{Shaw1993}. This concept of nonlinear modes describes the kind of families of solutions, for which all dependent variables and their time derivatives are functionally related to a single pair of one dependent variable and its time derivative. The Shaw and Pierre definition of nonlinear modes is less general than merely requiring periodicity. It is more specific regarding the properties of solutions, as it contains all oscillations evolving in a two-dimensional submanifold of the $2n$-dimensional phase space, according to (ii). This mode definition covers for example also modal solutions for non-conservative systems. The definition of modes for conservative nonlinear systems proposed by \emph{Rosenberg} \cite{Rosenberg1966} 
defines the motions corresponding to a mode as \enquote{vibration-in-unison}, which means that all dependent variables reach their extrema and cross zero simultaneously. In other words, the dependent variables evolve on a curve, according to (iii). Rosenberg modes have been investigated respectively detected for nonlinear systems with Euclidean metrics so far, see, \eg, \cite{Rosenberg1966,Caughey1990,Vakakis1996,Georgiades2009}, for which the metric tensor $g$ is constant but $\nnabla f$ is nonlinear. A sub-class of the general Rosenberg modes is given by cases, where the curve is a straight line \cite{Rosenberg1966}. Thereby, the time evolution of all dependent variables is geometrically similar and therefore the modes are called \emph{similar nonlinear normal modes}. It becomes obvious from the  examples treated in \cite{Rosenberg1966,Georgiades2009}
 that only the similar nonlinear normal modes satisfy property (iv), of being invariant \wrt the energy (or, equivalently, the initial velocity). 
 
 In the present paper it will be shown that the velocity invariance property (iv) of straight modal lines is related to the Euclidean metric of the considered examples and this concept will be generalized for Riemannian manifolds.    Therefore, the novel definition of a \emph{strict normal mode} for conservative nonlinear systems will be introduced as curve in configuration space, which is invariant for any initial velocity along the curve. The paper states and proves necessary and sufficient conditions for the existence of \emph{strict normal modes} for nonlinear systems with Riemannian metric (containing as particular case, of course, the Euclidean metric). The main contribution is thus to completely characterize such nonlinear modes for the general class of
 conservative nonlinear systems \eqref{eq:differential-equations} evolving on Riemannian manifolds. Moreover, a constructive example of such a mode is provided for a double pendulum subject to an elastic potential field.

The present work is motivated by the study of fast locomotion both in biological and robotic systems. Such systems are highly nonlinear due to the highly coupled multi-body dynamics and the nonlinear compliance of the actuation system (be it muscles and tendons or gearboxes and cable drives). It is well-know from literature that for example running for a large variety of animals and for humans can be very well approximated by a template dynamics of low order, for example the so-called spring-loaded inverted pendulum \cite{Holmes2006,Maus2010}. The central hypothesis motivating the research presented in this paper is that the low-dimensional motion templates are strongly related to the intrinsic dynamic properties of the considered systems. We are convinced that a well developed theory of nonlinear oscillation modes is an essential tool for the understanding of locomotion in nature and for its technological replication.     

\section{Main Result}

\begin{definition}[Strict normal Mode]
	Let $\mathcal C := \gamma(\mathcal N)\subset\mathcal M$ be a one-dimensional smooth submanifold of $\mathcal M$, a curve, defined by the smooth map $\gamma : \mathcal N\to\mathcal M$ between the interval  $\mathcal N \subset 
	\mathbb{R}$ and the smooth $n$-manifold $\mathcal M$.
	$\mathcal C$ is referred to as a \textbf{strict normal mode}, if its associated tangent bundle $T_{\;}\mathcal C$ constitutes an invariant set of the differential equations \eqref{eq:differential-equations}. 
\end{definition}

\begin{theorem}
\label{theorem:the_only}
	$\mathcal C$ is a \textbf{strict normal mode} of the differential equations \eqref{eq:differential-equations}, if and only if
	\begin{enumerate}[label=(\alph*)]
		\item $\mathcal C$ is a geodesic (or autoparallel line w.r.t. the Levi-Civita connection)
		and
		\item the gradient vector $\nnabla f$ of the scalar function $f$ on $\mathcal C$ is tangential to $\mathcal C$, \ie, $\left(\nnabla f\right)_p\in T_p\mathcal C$, $\forall p\in\mathcal C$.
	\end{enumerate}
\end{theorem}

The following straight-forward property of geodesics will be used at two stages in the proof of the theorem.
\begin{lemma*}
	Let $\bs{\dot{\gamma}_1},\bs{\dot{\gamma}_2} : \mathcal C\to T_p\mathcal C$ be non-zero vector fields tangent to a curve $\mathcal C$. Then, the covariant derivative of $\bs{\dot{\gamma}_1}$ \wrt $\bs{\dot{\gamma}_2}$ either vanishes or is again tangent to $\mathcal C$, $\left(\nnabla_{\bs{\dot{\gamma}_2}}\bs{\dot{\gamma}_1}\right)\in T_p\mathcal C$, if and only if $\mathcal C$ is a geodesic. 
\end{lemma*}

\begin{proof*}
	Let $\ww : \mathcal C\to T_p\mathcal C$, $\left<\ww, \ww\right>=1$, be a unit vector field tangent to $\mathcal C$, i.e. the tangent vector field arising from the arc length parametrization of the curve.  Further, let $\alpha,\beta : \mathcal C\to\RRR$ be non-zero scalar functions on $\mathcal C$ such that $\bs{\dot{\gamma}_1} = \alpha\ww$ and $\bs{\dot{\gamma}_2} = \beta\ww$. Then,
	\begin{align}
	\nnabla_{\bs{\dot{\gamma}_2}}\bs{\dot{\gamma}_1} = \nnabla_{\beta\ww}\left(\alpha\ww\right) = \beta\left(\nabla_{\ww}\alpha\right)\ww + \alpha\beta\nnabla_{\ww}\ww\,. 
	\end{align}
	$\left(\nabla_{\ww}\alpha\right)$ is a scalar function on $\mathcal C$, while $\nnabla_{\ww}\ww=\nnull$, if and only if $\mathcal C$ is a geodesic. 
\end{proof*}
Basically, the lemma recalls that for any time evolution of the considered system, the covariant derivative is tangent to the geodesic, the same way as for any motion along a straight line in Euclidean space the acceleration is a vector along that line.  Furthermore, the converse also holds: if for any system evolution the covariant derivative of its velocity vector field is tangent to the curve, then the curve is a geodesic. 

\begin{proof*}[Proof of theorem: Sufficiency]
From $\left(\nnabla f\right)_p\in T_p\mathcal C$, $\forall p\in\mathcal C$ it follows that $\nnabla f$ on $\mathcal C$ can be expressed as
	\begin{align}
	\left(\nnabla f\right)_p = \alpha \ww\,,
	\label{eq:tangent-gradient-vector}
	\end{align}
	where $\alpha : \mathcal C\to\RRR$ is a scalar function, and $\ww : \mathcal C\to T_p\mathcal C$, $\left<\ww, \ww\right>=1$, is a unit vector field tangent to $\mathcal C$. Accordingly, the differential equations \eqref{eq:differential-equations} on $T_{\;}\mathcal C$ satisfy
	\begin{align}
	\nnabla_{\bdgamma}\bdgamma = \alpha\ww \,.
	\label{eq:dgl-on-geodesic}
	\end{align}
	Now, choose the ansatz  $\bdgamma = \beta\ww$ with $\beta : \mathcal C\to\RRR$ a scalar function as a solution for \eqref{eq:dgl-on-geodesic}. Then, 
	\begin{align}
	\nnabla_{\bdgamma}\bdgamma 
	= \beta\left(\nabla_{\ww}\beta\right)\ww\,, \label{eq:ode-beta-times-w}
	\end{align}
	according to the above lemma. From \eqref{eq:ode-beta-times-w} it follows that the solution of \eqref{eq:dgl-on-geodesic} is $\bdgamma = \beta\ww$, if the differential equation
	\begin{align}
	\beta\nabla_{\ww}\beta = \alpha
	\label{eq:ode-beta}
	\end{align}
	can be solved for $\beta$. Selecting local coordinate charts $(\mathcal U,\phi_{\mathcal U})$ and $(\mathcal V,\phi_{\mathcal V})$ for $\mathcal M$ and $\mathcal N$ with local coordinates $x\in\RRR^n$ and $s\in\RRR$, respectively, such that $x(s) := \phi_{\mathcal U}\left(\gamma\circ\phi_{\mathcal V}^{-1}\right) : \RRR\to\RRR^n$, $\alpha(s) = \alpha\circ\phi_{\mathcal V}\circ\gamma^{-1} : \RRR\to\RRR$, and $\beta(s) = \beta\circ\phi_{\mathcal V}\circ\gamma^{-1} : \RRR\to\RRR$, \eqref{eq:ode-beta} takes the form
	\begin{align}
	\beta(s)\d\beta(s) = \alpha(s)\d s \,,
	\end{align}
	for which a solution always exists and is
	\begin{align}
	\frac{1}{2}\beta^2(s) + c = \int_0^s\alpha(\sigma)\d\sigma \,,
	\end{align}
	where $c$ is a constant of integration. This proves that \eqref{eq:dgl-on-geodesic} has always as a solution a vector field, which is tangent to the geodesic, and therefore sufficiency can be concluded.
\end{proof*}
\begin{proof*}[Proof of theorem: Necessity]
	Recall that, since $\mathcal C$ is an embedded submanifold of $\mathcal M$, for any $p\in\mathcal C$,
	\begin{align}
	T_p\mathcal M = T_p\mathcal C \oplus \left(T_p\mathcal C\right)^\perp\,,
	\end{align}
	that is any vector $\XX\in T_p\mathcal M$ may be written as the sum of a vector $\XX^\top\in T_p\mathcal C$ and a normal vector $\XX^\perp := \XX - \XX^\top$.
	
	Assume, for the sake of contradiction that $\mathcal C$ is not a geodesic, \ie,
	\begin{align}
	\left(\nnabla_{\bdgamma}\bdgamma\right)_p = \left(\nnabla_{\bdgamma}\bdgamma\right)_p^\top + \left(\nnabla_{\bdgamma}\bdgamma\right)_p^\perp\,,\quad\forall (p,\bdgamma)\in T_{\;}\mathcal C \,,
	\end{align} 
	with nonzero normal component according to the above lemma. 
	Then, satisfying the differential equations \eqref{eq:differential-equations} on $T_{\;}\mathcal C$ requires that
	\begin{align}
	\left(\nnabla_{\bdgamma}\bdgamma\right)_p^\perp = (\nnabla f)_p^\perp\,,
	\label{eq:acceleration}
	\end{align}
	where $(\nnabla f)_p^\perp$ is the normal component of the potential gradient.
	However,
	\begin{align}
	\left(\nnabla_{c\bdgamma}c\bdgamma\right)_p^\perp = c^2\left(\nnabla_{\bdgamma}\bdgamma\right)_p^\perp\,,
	\label{eq:scaled-vector-field}
	\end{align}
	for any tangent vector field scaled by a constant $c$. According to the definition of strict normal modes, \eqref{eq:differential-equations} has to be satisfied also for the scaled tangent vector field, which implies 
	\begin{align}
	c^2\left(\nnabla_{\bdgamma}\bdgamma\right)_p^\perp = (\nnabla f)_p^\perp\,.
	\end{align}
	This contradicts \eqref{eq:acceleration}, since $f$ is independent of the velocity and therefore $(\nnabla f)_p^\perp$ is independent of $c$. We therefore conclude that $\mathcal C$ is a geodesic. Thus $\left(\nnabla_{\bdgamma}\bdgamma\right)_p^\perp$ is zero implying by \eqref{eq:acceleration} that also $(\nnabla f)_p^\perp$ must be zero, i.e. $\nnabla f$ on $\mathcal C$ is tangent to $\mathcal C$.
\end{proof*}

An immediate consequence of the theorem is 
\begin{corollar}
In systems with Euclidean metric, strict normal modes are straight lines.
\end{corollar}
This explains why in most of the literature \cite{Rosenberg1966,Shaw1993,Vakakis1996,Georgiades2009}, which considered normal modes of systems consisting of point masses connected by nonlinear springs, only lines happened to be invariant with respect to the initial velocity (or equivalently, w.r.t. the initial energy level). To our knowledge, our previous publication \cite{Lakatos2017} presented the first example of a strict nonlinear normal mode for a system with non-Euclidean metric.   

 The above theorem does not make any statement regarding periodicity of motion, but only about the invariance of the curve. Thus, a remark on the structure of the potential function $f$ should be made.
 
\begin{corollar}
\label{corollar_periodicity}
If the scalar function $f$ is negative definite on the curve $\mathcal C$, having an equilibrium point $p$ on the curve, then the system will perform periodic oscillations around this point.
\end{corollar}
Periodicity can be concluded for this one-dimensional problem based on the Poincar\'e-Bendixson theorem \cite{Strogatz2015}. 
Under the additional conditions of Corollary \ref{corollar_periodicity}, the strict normal modes fulfil all criteria (i)-(iv).

\section{Examples of Strict Normal Modes}
\begin{figure}[htb]
	\includegraphics[width=6.0cm]{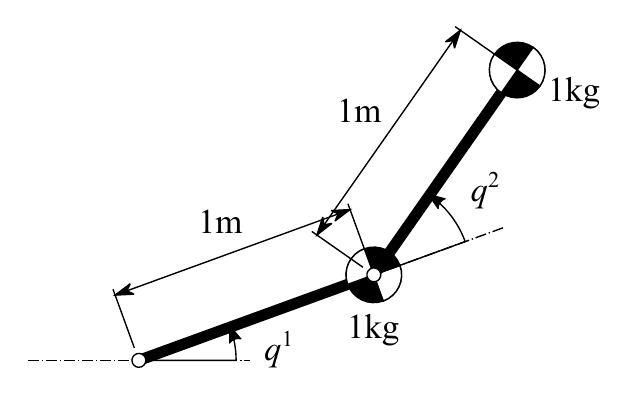}
	\caption{Inertia model of the double pendulum considered as example for a system with Riemannian metric tensor.}
	\label{fig:double_pendulum}
\end{figure}

A double pendulum is considered as an example of a two-dimensional, nonlinear system, having a non-Euclidean metric tensor. First, a numerical analysis for a basic potential field is given, to visualize the various mode properties mentioned in the introduction.  Thereafter, the above theorem is applied to render an arbitrary geodesic curve into a strict (nonlinear) normal mode of oscillation. In contrast to the theoretical result at hand, (which is formulated in a coordinate-free way) coordinates will be introduced here to solve the specific problem. \\
Consider a planar double pendulum, \ie, two regular pendulums hinged to each other, with unit lengths and unit point masses at the tip of each pendulum as shown in Fig.~\ref{fig:double_pendulum}.

\begin{figure*}[htb]
	\includegraphics[width=\textwidth]{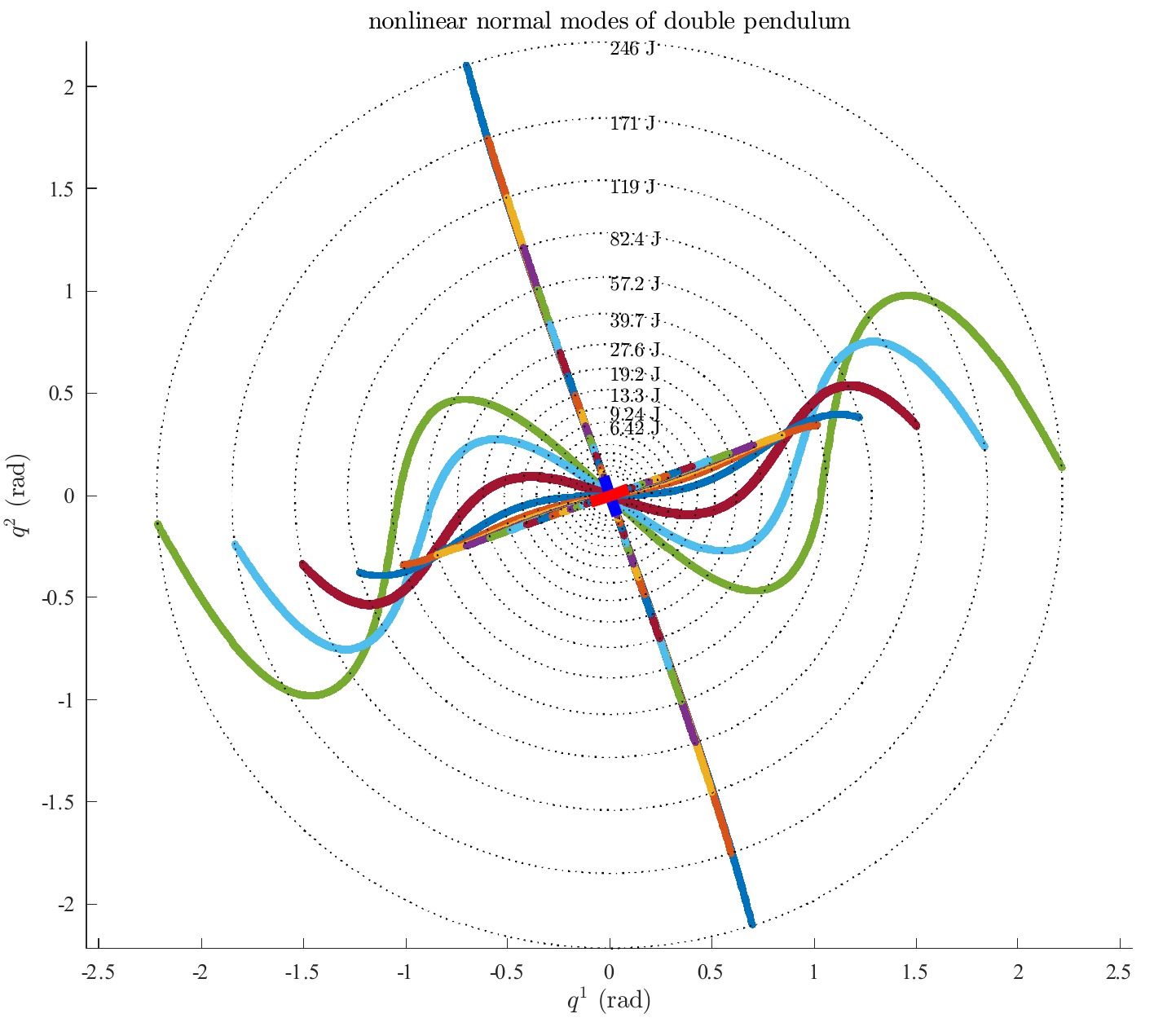}
	\caption{Nonlinear normal modes of the planar double pendulum with circular potential field. The thick, red and blue lines represent the eigenvectors of the system linearization at the equilibrium point $(q^1=q^2=0, \dot q^1=\dot q^2=0)$. The dotted circles indicate energy levels of the system. For each energy level, the corresponding two modes are displayed. Therefore, the system is simulated for 200s, corresponding to approx. 80-90 periods. The modes are found by optimization of initial configurations on the equipotential line, the cost function being given by a periodicity measure (autocorrelation). As the velocity of the system is zero at the ends of the normal modes, they end on the equipotential lines \cite{Rosenberg1966}.  While one mode is to a good approximation strict, i.e. the curve is invariant with respect to energy, the other mode strongly deforms when energy increases.}
	\label{fig:mode_plots}
\end{figure*}
Let us introduce coordinates $q = (q^1,q^2)\in\RRR^2$, where $q^1$ measures the absolute angle of the first pendulum, and $q^2$ measures the relative angle between the first and second pendulum. This choice of coordinates results in an inertia tensor with components
\begin{align}
\begin{split}
\tensors{g}{}{11} &= 3+2\cos q^2 \,, \\
\tensors{g}{}{12} &= \tensors{g}{}{21} = 1+\cos q^2 \,, \\
\tensors{g}{}{22} &= 1 \,.
\end{split}
\label{eq:double-pendulum-inertia}
\end{align}
\subsection{Numerical Analysis for a Simple Potential Function}
Fig.~\ref{fig:mode_plots} visualizes the nonlinear normal modes of the system for the case that a linear spring with stiffness $k_0=100\,\mathrm{Nm/rad})$ acts on each joint. This means that the potential function has the form $f(q)=-\frac{1}{2}k_0((q^1)^2 + (q^2)^2)$, i.e., the equipotential lines are circles. The eigenmodes of the linearized system are displayed by thick (blue and red) lines. It is known that for small amplitudes of nonlinear systems there exist at least as many periodic solutions as the number of modes of the linearized system. This has been shown by Lyapunov \cite{Lyapunov1884} for the special case of linear eigenvalues which are not rationale multiples and by \cite{Weinstein1973} for a more general case. As can be seen from Fig.~\ref{fig:mode_plots}, the eigenmodes of the nonlinear system are (not surprisingly) very similar to the linear ones for low energy levels. More intriguing is that periodic solutions exist also for large amplitudes. They continuously deform into nonlinear curves as the energy level of the oscillation increases. \\
A qualitative difference between the two nonlinear normal modes can be seen: while one mode strongly deforms as energy increases, the second one stay approximately on the same curve, with only the amplitude being increased.  According to the terminology introduced in this paper, the first mode would be a \emph{energy dependent normal mode}, while the second mode closely resembles\footnote{We cautiously say "closely resembles" because the numerical analysis does not constitute a proof that this indeed is a strict mode in the considered energy range.} a \emph{strict mode}. Note that the energy dependent mode does not correspond entirely to the nonlinear normal mode definition of Rosenberg \cite{Rosenberg1966}: while $q^2$ can be indeed expressed as a function of $q^1$, the mapping is not injective. This implies that, while both velocities become zero on the equipotential line,  $\dot q^2$ has also other zeros. Definitely, further investigations need to be done to more deeply understand the nature and properties of energy-dependent modes on Riemannian manifolds.      

\begin{figure}[htb]
	\includegraphics[width=\columnwidth]{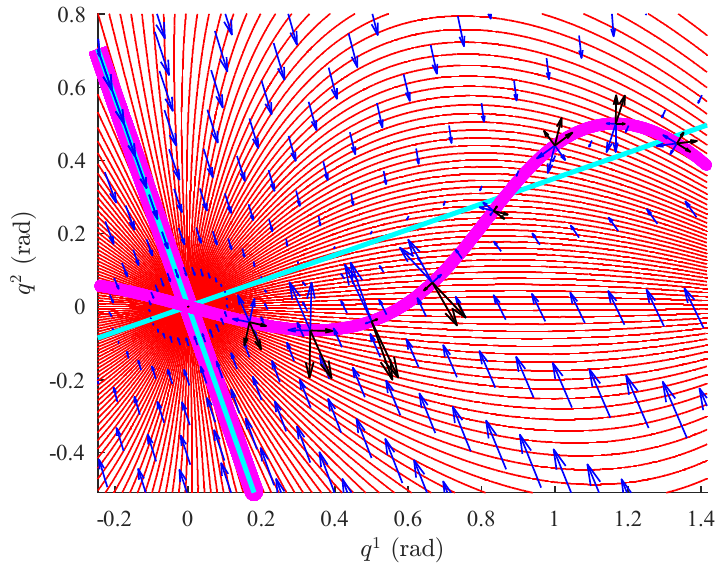}
	\caption{Visualization of the theorem statements for the planar double pendulum. In red, the geodesics starting from the equilibrium point are displayed. The modes of the linearized system are shown in cyan while the nonlinear modes are displayed by thick magenta lines. The gradient field of the potential (scaled by a constant factor for better visualization) is displayed by the blue arrow field.  One mode is strict to a good approximation: it evolves along a geodesic and the potential gradient is tangent to it. Although close to origin it is a straight line, corresponding also to the linear mode, for larger amplitudes it has slight deviations from the linear mode, 
    as can be recognized in Fig.~\ref{fig:mode_plots}. The second mode clearly does not fulfil the theorem conditions. Neither is it a geodesic curve nor is the potential gradient tangential to the mode. The potential gradient along with its tangential and normal decomposition are displayed in blue for various points on the mode. The covariant derivative along with its tangential and normal decomposition are displayed in black.}   
	\label{fig:geodesics_and_theorem}
\end{figure}
In order to visualize the statements of the theorem, Fig.~\ref{fig:geodesics_and_theorem} displays geodesics starting from the equilibrium point, the modes of the linearized system, the nonlinear modes, and the gradient field of the potential. One can observe also here that one mode fulfils to a good approximation the conditions of a strict mode: the gradient evolves tangentially to a geodesic. The modal curve remains therefore quasi invariant when increasing the amplitude.\\
For the second mode, which bends when energy (and thus velocity) increases, the gradient and the covariant derivative, along with their normal and tangential decomposition, are displayed at several points. As expected, this mode does not correspond to a geodesic. The covariant derivative and the gradient are not tangent to the modal curve.

\subsection{Construction of a Strict Mode by Potential Function Design}
\begin{figure}[htb]
\centering
	 \includegraphics[width=7.5cm]{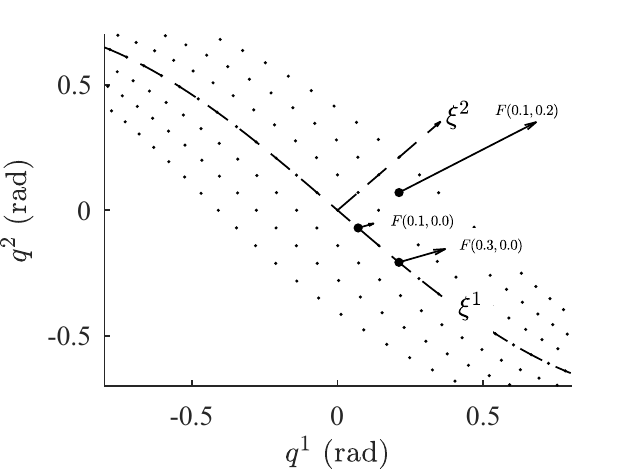}
	\caption{Grid with resolution $0.1$ of the local geodesic and transverse coordinates $\xi^1$ and $\xi^2$ are depicted, respectively. Additionally, forces at three points of the force field, satisfying condition (b) of the theorem, are shown.}
	\label{fig:force_field}
\end{figure}

Now let us apply theorem~\ref{theorem:the_only} to turn an arbitrary geodesic curve into a strict mode.
A geodesic corresponding to the above metric field $g$ can be expressed as a parametrized curve $q=\gamma(\xi^1)$, where $\xi^1\in\RRR$ is the coordinate on the curve and ${\gamma : \RRR \to \RRR^2}$. Let us introduce another coordinate $\xi^2\in\RRR$ in the direction perpendicular to the curve. This local coordinate system defines a local diffeomorphism ${h : \RRR^2\to\RRR^2}$ between $(\xi^1,\xi^2)$ and $(q^1,q^2)$, as shown in Fig.~\ref{fig:force_field},
\begin{align}
h(\xi) = \gamma(\xi^1) + \xi^2 e^{\perp}(\xi^1)
\label{eq:local-diffeomorphism}
\end{align}
with components of the normal basis vector,
\begin{align}
e^{\perp}(\xi^1) = 
\begin{pmatrix}
0 & -1 \\ 1 & 0
\end{pmatrix}
\td{\gamma(\xi^1)}{\xi^1} \,.
\end{align}
Note that the inverse of \eqref{eq:local-diffeomorphism} maps the geodesic curve to a straight line.
The goal is to construct a potential function $f$, which satisfies condition (b) of the above theorem. To obtain a nonlinear system displaying periodic orbits on the geodesic, $f$ is constructed to be negative definite (\wrt to a point on the geodesic). To this end, consider the components of a force field $\tensors{F}{}{i}(\xi^1,\xi^2)$ expressed in geodesic coordinates defined by \eqref{eq:local-diffeomorphism}. On the geodesic, \ie, $\forall\xi^1\in\RRR$ and $\xi^2=0$, the components of such a force field may take the form
\begin{align}
\tensors{F}{}{i}(\xi^1,\xi^2=0) = \alpha(\xi^1)\left.\pdx{h^j(\xi)}{\xi^i}\right|_{\xi^2=0}\tensors{g}{}{jk}(\gamma(\xi^1))\td{\gamma^k(\xi^1)}{\xi^1} \,.
\end{align}
Herein $\alpha : \RRR\to\RRR$ is a scalar function satisfying $\td{\alpha(\xi^1)}{\xi^1} < 0$, $\forall\xi^1\in\RRR$, $\td{\alpha(\xi^1)}{\xi^1} = 0$, $\xi^1=0$ (such that $f$ has its maximum at $\xi^1=0$). The force field $\tensors{F}{}{i}(\xi^1,\xi^2)$ can be derived from a potential function $f : \RRR^2\to\RRR$, if the integrability condition
\begin{align}
\pdx{\tensors{F}{}{1}}{\xi^2} = \pdx{\tensors{F}{}{2}}{\xi^1}
\label{eq:2dof_integrability}
\end{align}
is satisfied. This can be achieved by construction, \eg,
\begin{align}
\tensors{F}{}{1}(\xi^1,\xi^2) = \tensors{F}{}{1}(\xi^1,0) + \int_0^{\xi^2}\pdx{\tensors{F}{}{2}(\xi^1,0)}{\xi^1}\d s \,.
\end{align}
Negative definiteness of $f$ in its arguments (which is a requirement for the mechanical implementation of the potential by elasticities) can be ensured by choosing
\begin{align}
\tensors{F}{}{2}(\xi^1,\xi^2) = \tensors{F}{}{2}(\xi^1,0) + \int_0^{\xi^2}\beta(s)\d s \,,
\label{eq_positive-definiteness}
\end{align}
where
\begin{align}
\beta(\xi^2) < \inf_{\xi^1\in[-\epsilon;\epsilon]}\frac{\left(\pdx{\tensors{F}{}{2}(\xi^1,0)}{\xi^1}\right)^2}{\pdx{\tensors{F}{}{1}(\xi^1,\xi^2)}{\xi^1}} \,,
\end{align}
for a certain $\epsilon$-neighborhood of $\xi^1$.

\begin{figure}[htb]
\centering
	\includegraphics[width=7.5cm]{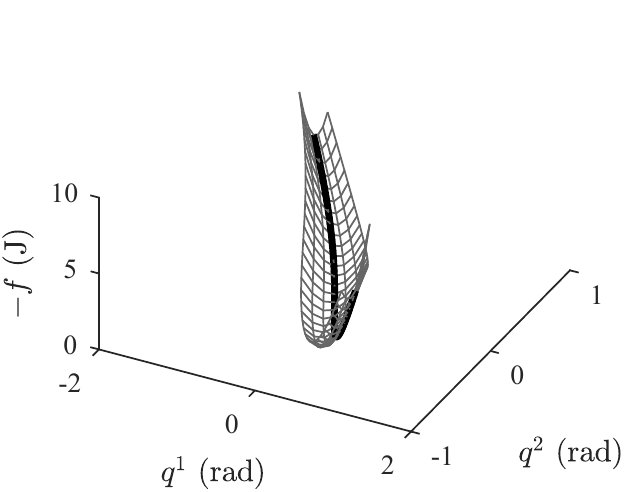}
	\caption{Potential function constructed to satisfy condition (b) of the theorem.}
	\label{fig:potential}
\end{figure}

\begin{figure}[htb]
\centering
	\includegraphics[width=7.5cm]{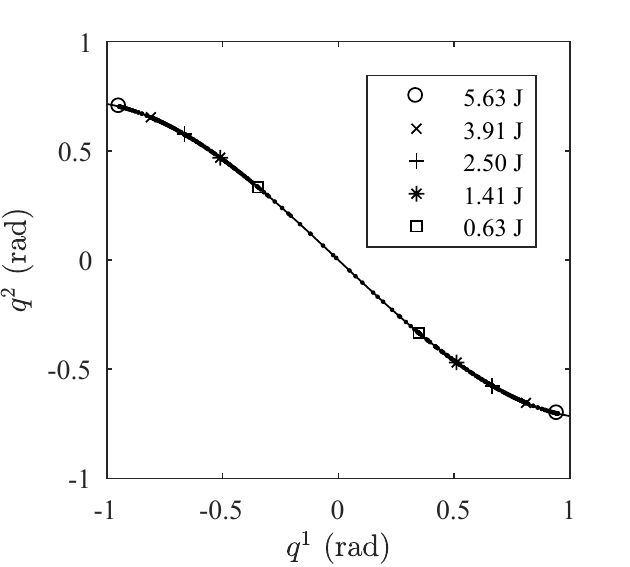}
	\caption{Geodesic curve (solid line). Numerical solution of the differential equations (dots). Point of maximum potential energy for each solution (markers).}
	\label{fig:configuration-space-evolution}
\end{figure}

\begin{figure}[htb]
\centering
	\includegraphics[width=7.5cm]{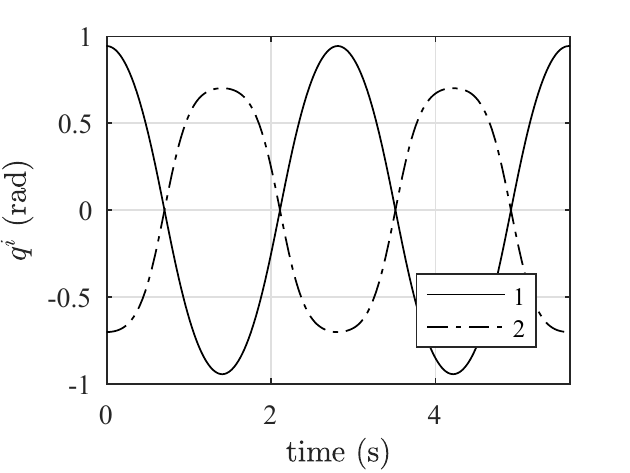}
	\caption{Time evolution of the pendulum angles $q^1$ and $q^2$ corresponding to oscillations at an energy level of $5.63\,\mathrm{J}$, obtained by numerical integration.}
	\label{fig:time-evolution}
\end{figure}

In Fig.~\ref{fig:potential} the potential function $-f(q^1,q^2)$  is depicted, where $\alpha =-5\xi^1$ and $\beta=-47.86=\t{const.}$. $f$ satisfies condition (b) of the theorem for a geodesic induced by the double pendulum inertia tensor \eqref{eq:double-pendulum-inertia}. The geodesic curve $\gamma(s)$ is obtained by solving the initial value problem ${(\nabla_{\dot \gamma}\dot \gamma)^i = 0}$, ${\gamma^1(0)=0}$, ${\gamma^2(0)=0}$, ${\dot\gamma^1(0)=\cos(-\pi/4)}$, ${\dot\gamma^1(0)=\sin(-\pi/4)}$. The numerical solutions of the differential equations \eqref{eq:differential-equations} (characterized by the above $f$ and $g$) for initial conditions at different energy levels evolve on the geodesic, as shown in Fig.~\ref{fig:configuration-space-evolution}. This validates that the geodesic is a strict normal mode according to the above definition. In Fig.~\ref{fig:time-evolution} the time evolution of the physical coordinates (pendulum angles) $q^1$ and $q^2$ corresponding to oscillations at an energy level of $5.63\,\mathrm{J}$ (obtained by numerical integration) are shown. Herein, unison oscillations of $q^1$ and $q^2$ can be observed, which are in accordance of the definition of \emph{normal modes} introduced by Rosenberg \cite{Rosenberg1966}, however obtained under way more general conditions than the so-called \emph{similar normal modes} (modal lines) in \cite{Rosenberg1966}. 

\section{Conclusions}
As discussed in this paper, there is a rich structure of periodic solutions of invariant low-dimensional manifolds even for highly nonlinear systems characterized by a Riemannian metric and a non-quadratic potential field. The paper demonstrates that invariant curves in configuration space can only be geodesics and require a special alignment of the potential field w.r.t. the geodesic. The paper provides a constructive way of designing technical systems having this dimensionality reduction property. But also without this special design procedure, strict nonlinear normal modes seem not to be an exception, as indicated by the presented numerical example. While in \cite{Lakatos2017} we presented, to our knowledge, the first example of a system with Riemannian metric exhibiting a strict mode, the differential geometric perspective allows the formulation of the very general theorem and of the related strict mode design procedure.\\   
The presented elastic double pendulum is an extremely simplified model of a biological limb. From biomechanics and neuroscience perspective, it is therefore interesting to ask, if biological bodies (and the related neural control) developed through evolution and can further adapt individually (by bone and muscle growth)  such that the conditions of the theorem are fulfilled.  As the results are formulated in a quite general way, the applications might reach however far beyond the above motivating examples.

\begin{ack}                           
This work has been partially funded by the ERC Advanced Grants M-RUNNERS and PORTWINGS 
\end{ack}
\appendix
\section*{Appendix}

The notions of geodesic, shortest line, parallel transport and straight line on manifolds are summarized for convenience in the following, see for example \cite{Frankel2003}, pp.232-290.

\begin{definition}[Shortest Curve]
	If the manifold $\mathcal M$ has a Riemannian structure $g$, then the curve $\mathcal C \subset \mathcal M$ is said to be the \textbf{shortest} connecting the points $a, \,b \in \mathcal M$,
	if the curve is what is said a \textit{geodesic} for $g$, which means that it is an extremal of the length integral 
	\begin{equation}
	    L(\delta)=\int_a^b g( \dot \ggamma(t), \dot \ggamma(t)) \mathrm{d}t
	\end{equation}
	among variations $\mathcal C_\delta$ of the curve, where $t$ is \emph{any} parametrization of the curves and $\dot \ggamma(t)$ is the derivative with respect to $t$, i.e., a tangent vector field along the curve.
\end{definition}

\begin{definition}[Straight Curve]
	If the manifold $\mathcal M$ has a connection $\nnabla$, the curve $\mathcal C$ is said to be \textbf{straight}, if the curve is what is called \textit{autoparallel} for $\nnabla$, which means that 
	 $\left(\nnabla_{\bdgamma}\bdgamma\right)_p = \nnull$, $\forall p\in\mathcal C$ and for any unit tangent vector $\bdgamma : \mathcal C\to T_p\mathcal C$, (satisfying ${\left<\bdgamma, \bdgamma\right> = \t{const.}}$). 
\end{definition}
A unit tangent vector $\dot \ggamma(t)$ is obtained for example if the curve $\mathcal C$ is described by $\ggamma(t)$ with $t$ being arc length parameterization.
From the previous definitions, it is clear that if we define a Levi-Civita connection based on a Riemannian metric ($\nnabla g =0$), we recover the Euclidean notion that straight lines are the shortest lines between two points. It is important not to mix the concept of straight curves with the concept of curvature (i.e. the curvature being zero). In fact, not only on a Euclidean manifold, but also in a curved space one can talk about straight and short lines. The difference in a curved space appears when considering what is called the parallel transport of sections along general curves. In such cases, for a curved space, the result of the transport of a general section will depends on the chosen line. 

\bibliographystyle{plain}
\bibliography{refLakatos}

\begin{thebibliography}{10}

\bibitem{Caughey1990}
T.K. Caughey, A.~Vakakis, and J.M. Sivo.
\newblock Analytical study of similar normal modes and their bifurcations in a
  class of strongly non-linear systems.
\newblock {\em Int. J. of Non-Linear Mechanics}, 25(5):521--533, 1990.

\bibitem{Frankel2003}
Th. Frankel.
\newblock {\em {The Geometry of Physics: An Introduction}}.
\newblock Cambridge University Press, 2 edition, 2003.

\bibitem{Georgiades2009}
F.~Georgiades, M.~Peeters, G.~Kerschen, J.C. Golinval, and M.~Ruzzene.
\newblock Modal analysis of a nonlinear periodic structure with cyclic
  symmetry.
\newblock {\em AIAA J.}, 47(4):1014--1025, 2009.

\bibitem{Holmes2006}
P.~Holmes, R.J. Full, D.~Koditschek, and J.~Guckenheimer.
\newblock The dynamics of legged locomotion: Models, analyses and challenges.
\newblock {\em SIAM Review}, 48(2):207--304, 2006.

\bibitem{Lakatos2017}
D.~Lakatos, W.~Friedl, and A.~Albu-Sch\"affer.
\newblock Eigenmodes of nonlinear dynamics: Definition, existence, and
  embodiment into legged robots with elastic elements.
\newblock {\em Robotics and Automation Letters}, 2(2):1062--1069, Apr. 2017.

\bibitem{Lyapunov1884}
A.~M. Lyapunov.
\newblock {\em The General Problem of Stability of Motion}.
\newblock Taylor and Francis, Translated and Edited by A. T. Fuller: 1992,
  original publication in Russian: 1884.

\bibitem{Maus2010}
H.-M. Maus, S.W. Lipfert, M.~Gross, J.~Rummel, and A.~Seyfarth.
\newblock Upright human gait did not provide a major mechanical challenge for
  our ancestors.
\newblock {\em Nature communications}, 1:70, 2010.

\bibitem{Rand1971}
R.H. Rand.
\newblock Nonlinear normal modes in two-degree-of-freedom systems.
\newblock {\em J. of Applied Mechanics}, 38(2):561--561, 1971.

\bibitem{Rand1974}
R.H. Rand.
\newblock A direct method for non-linear normal modes.
\newblock {\em Int. J. of Non-linear Mechanics}, 9(5):363--368, 1974.

\bibitem{Rosenberg1966}
R.M. Rosenberg.
\newblock On nonlinear vibrations of systems with many degrees of freedom.
\newblock {\em Advances in Applied Mechanics}, 9:155--242, 1966.

\bibitem{Shaw1993}
S.W. Shaw and C.~Pierre.
\newblock Normal modes for non-linear vibratory systems.
\newblock {\em J. of Sound and Vibration}, 164(1):85--124, 1993.

\bibitem{Strogatz2015}
S.~Strogatz.
\newblock {\em Nonlinear Dynamics and Chaos}.
\newblock Perseus Books, 2015.

\bibitem{Vakakis1996}
A.F. Vakakis, L.I. Manevitch, Y.V. Mikhlin, V.N. Pilipchuk, and A.~A. Zevin.
\newblock {\em Normal Modes and Localization in Nonlinear Systems}.
\newblock Wiley, 1996.

\bibitem{Weinstein1973}
A.~Weinstein.
\newblock Lagrangian submanifolds and {Hamiltonian} systems.
\newblock {\em Annals of Mathematics, Second Series, Published by Mathematics
  Department, Princeton University, Stable URL:
  https://www.jstor.org/stable/1970911}, 98(3):377--410, 1973.

\bibitem{Zhang2005}
X.~Zhang.
\newblock Geodesics, nonlinear normal modes of conservative vibratory systems
  and decomposition method.
\newblock {\em J. of Sound and Vibration}, 282(3):971 -- 989, 2005.

\end{thebibliography}

\end{document}